\documentclass[prd, twocolumn, nofootinbib, floatfix]{revtex4-1}

\usepackage{amsmath}
\usepackage{graphicx}
\usepackage{dcolumn}
\usepackage{bm}
\usepackage{enumerate} 
\usepackage{epsfig}
\usepackage{amssymb,latexsym,mathrsfs}
\usepackage{graphicx}
\usepackage{color}
\usepackage{hyperref}

\hypersetup{
    colorlinks=true,
    linkcolor=red,
    citecolor=blue,
} 

\newcommand{\be}{\begin{equation}}
\newcommand{\ee}{\end{equation}}
\newcommand{\bes}{\begin{equation*}}
\newcommand{\ees}{\end{equation*}}
\newcommand{\beq}{\begin{equation}}
\newcommand{\eeq}{\end{equation}}
\newcommand{\bs}{\begin{split}} 
\newcommand{\bea}{\begin{eqnarray}}
\newcommand{\eea}{\end{eqnarray}}
\newcommand{\beqa}{\begin{eqnarray}}
\newcommand{\eeqa}{\end{eqnarray}}

\newcommand{\om}{\Omega_m}

\newcommand{\wowa}{$w_0$--$w_a$}

\begin{document}

\title{Quintessence's Last Stand?} 
\author{Eric V.\ Linder} 
\affiliation{Berkeley Center for Cosmological Physics \& Berkeley Lab, 
University of California, Berkeley, CA 94720, USA} 

\begin{abstract}
Current cosmological data puts increasing pressure on models of dark 
energy in the freezing class, e.g.\ early dark energy or those with 
equation of state $w$ substantially different from $-1$. We investigate 
to what extent data will distinguish the thawing class of quintessence 
from a cosmological 
constant. Since thawing dark energy deviates from $w=-1$ only at late 
times, we find that deviations $1+w\lesssim0.1$ are difficult to see even 
with next generation measurements; however, modest redshift drift data 
can improve the sensitivity by a factor of two. Furthermore, technical 
naturalness prefers specific thawing models. 
\end{abstract}

\date{\today} 

\maketitle

\section{Introduction} 

Cosmic acceleration indicates crucial new physics exists outside the 
standard model of particle physics and cosmology. Cosmological data 
imply the expansion history of the universe acts like the matter component 
is supplemented with a fluid of strongly negative effective pressure: 
dark energy with an effective equation of state, or pressure to energy 
density, ratio $w\approx-1$. One possibility is the cosmological constant, 
with $w=-1$ for all times, or redshifts $z$, and no spatial perturbations. 
For data sensitive to only the background cosmic expansion, or Hubble 
parameter $H(z)$, the effects are fully described by $w(z)$, even if there 
is no physical dark energy fluid but rather a modification of the form of 
the action. 

Cosmic microwave background (CMB) measurements impose two stringent 
constraints on the dark energy. From the distance to last scattering, the 
energy density weighted time average of $w(z)$ must be close to $-1$ 
\cite{planckw}. 
From the structure of the CMB temperature power spectrum, the dark energy 
density around recombination ($z\approx10^3$) is less than about 1\% of 
the critical energy density \cite{planckw,planckede,hojls}. 
Late time observations of the cosmic expansion, 
such as supernovae or baryon acoustic oscillation distances, also indicate 
that $w\approx-1$ at recent times \cite{betoule,bossw}. 
These results disfavor the first quintessence 
models, tracker models with insensitivity to initial conditions, but moreover 
most of the freezing region of dark energy dynamics (Ref.~\cite{caldlin} 
showed that the effective dark energy field either began dominated by 
Hubble friction -- thawing fields -- or by the steepness of the potential -- 
freezing fields, unless fine tuned). 

The remaining dark energy behavior favored by data is the thawing class. 
While thawing fields do not have the trackers' insensitivity to initial 
conditions, some of them do have the attractive quality of technical 
naturalness -- protection against quantum radiative corrections. 
Since thawing models only deviate from $w\approx-1$ at late times, they are 
consistent with the data but are also harder to distinguish from a static 
cosmological constant, despite having very different physics implications. 
Given the current state of data, and the advancing plans for next generation 
experiments, we assess whether continued consistency of the data with a 
cosmological constant $\Lambda$ will enable all quintessence models to be 
disfavored, or at least relegated to a region fine tuned to be close to 
$\Lambda$. 

Beyond quintessence, cosmic acceleration models with deviations in growth 
of structure relative to their expansion behavior, such as from modified 
gravity theories, or clustered or coupled dark energy, can also be constrained 
by growth measurements. However within quintessence -- canonical, minimally 
coupled dark energy -- no new signatures arise within growth. Thus the 
question of whether we will be able to distinguish thawing dark energy, and 
hence most of the remaining viable phase space of quintessence, from 
$\Lambda$ will focus on the expansion history. 

In Sec.~\ref{sec:thaw} we discuss methods of treating the thawing class of 
dark energy as a whole, drawing a clear distinction with inappropriate 
slow roll assumptions. We examine the leverage of future data 
in separating thawing quintessence from a cosmological constant in 
Sec.~\ref{sec:fit}, highlighting the role of potential redshift drift 
measurements. We explore some particle physics implications of fields 
nearly indistinguishable from $\Lambda$, and the virtues of technical 
naturalness, in Sec.~\ref{sec:field} and conclude 
in Sec.~\ref{sec:concl}.

\section{Thawing Quintessence} \label{sec:thaw} 

The class of thawing quintessence includes common potentials such as 
monomials $V(\phi)\sim\phi^n$ and pseudo-Nambu Goldstone boson (PNGB, or 
axion) fields with $V(\phi)\sim [1+\cos(\phi/f)]$, where $f$ is the symmetry 
breaking energy scale. Two members of this class are of particular interest 
since they have shift symmetry protecting against high energy radiative 
corrections. One is the PNGB case, long studied in terms of both natural 
inflation \cite{natinf} and dark energy \cite{frie95}. The other is the linear 
potential \cite{linpot,doom}. Recently the linear potential has garnered 
renewed interest in connection with both inflation \cite{conformal} and 
modified gravity \cite{sequester}. 

We are interested in constraining the thawing class as a whole, rather than 
individual models. To do this efficiently it is useful to parametrize its 
behavior. The standard $w_0$--$w_a$ form for the dark energy equation of 
state, $w(a)=w_0+w_a(1-a)$, where $a=1/(1+z)$ is the scale factor, works 
for a broad variety of dark energy models, 
beyond just the thawing class. Indeed it was derived from exact solutions 
of the field dynamics \cite{lin03} (and is completely unrelated to a linear 
or Taylor expansion). The $w_0$--$w_a$ form matches exact solutions of the 
observables -- distances and Hubble expansion $H$ -- to better than 0.1\% 
for a broad range of currently allowed models \cite{calde}. However, because 
it involves two parameters, constraints from even next generation data have 
difficulty distinguishing thawing models with $w_0\lesssim-0.9$ from a 
cosmological constant at 95\% confidence level, due to covariances between 
parameters. 

Since we focus our interest on the thawing class, we could use more 
specialized parametrizations, such as 
\cite{critt,alg,schersen,scherrer,chiba,bond}. 
We emphasize though that from an observational perspective the $w_0$--$w_a$ 
parametrization with its 0.1\% matching is wholly sufficient. To obtain 
tighter model constraints, we would need a one parameter form. We cannot 
rely on the usual inflationary slow roll field approximation 
since even thawing dark energy does not slow roll, in the sense that all 
terms in the Klein-Gordon equation for its dynamics are comparable 
\cite{paths}. For example, during the vast majority of e-folds when the 
field is frozen (e.g.\ during the matter dominated era), the terms are in 
the ratio of $1:2:-3$. Note also that a 
small field assumption is somewhat uncertain: while many thawing fields roll 
a distance $\Delta\phi\approx0.24\,M_{\rm Pl}$ to get to $w_0=-0.9$ today, 
they start at several $M_{\rm Pl}$ from their minima. 

We can take two approaches to adopting a one parameter thawing equation of 
state. Ref.~\cite{calde} showed that not only was \wowa\ an excellent 
approximation to exact solutions of the field evolution, and to observational 
quantities, but it could also be derived as a calibration relation for classes 
of dark energy dynamics. By appropriate choice of $w_0$ and $w_a$, the 
spread of evolution in the $w$--$w'$ plane calibrated into narrow tracks for 
different dark energy physics. In particular, a broad range of thawing models 
could all be tightly fit by a constrained $w_a$ form: 
\be 
w_a\approx -1.58\,(1+w_0) \ . \label{eq:wathaw} 
\ee 
This combines all the virtues of the general \wowa\ form (e.g.\ $0.1\%$ 
distance reconstruction) with the leverage of a one parameter fit. 

While this fits the expansion observables (distances and Hubble parameter) 
superbly, an alternate approach is to follow the thawing physics more 
closely. For example, thawing models have $w=-1$ at high redshift. This is 
not a problem for the form Eq.~(\ref{eq:wathaw}) since thawing dark energy 
density fades quickly into the past and so using $w(z\gg1)=w_0+w_a\ne-1$ 
has negligible effect. (Recall that even accounting for this the distance to 
high redshifts is accurately 
matched to better than 0.1\%.) If we felt more comfortable approximating 
$w(z)$, despite it not being an observable, then the algebraic thawing form 
of Ref.~\cite{alg} works quite well: 
\be 
1+w(a)=(1+w_0)\,a^p\,\left(\frac{1+b}{1+ba^{-3}}\right)^{1-p/3} \ , 
\label{eq:algorig} 
\ee 
where the constant $b=0.5$. This is 
derived from the physics of how the field evolves upon leaving the matter 
dominated era and automatically has the correct high redshift behavior. 

We can turn it into a one parameter form by fixing $p=1$, so the final 
algebraic form we will use is 
\be 
1+w(a)=(1+w_0)\,a^3\,\left(\frac{3}{1+2a^3}\right)^{2/3} \ . \label{eq:alg} 
\ee 
Note that of course it can fit a cosmological constant as well, with 
$w_0=-1$. Both the (general) algebraic thawed and the general \wowa\ 
forms have the advantage that the Hubble parameter $H(z)$ is analytic. 

\begin{figure}[htbp!]
\includegraphics[width=\columnwidth]{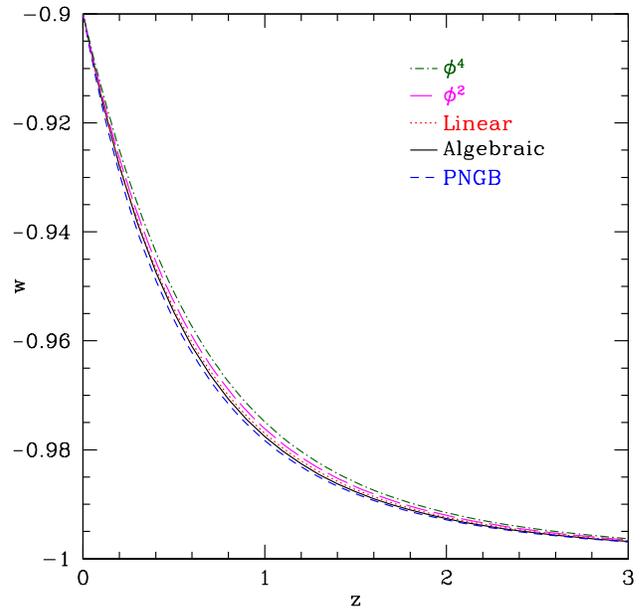} 
\caption{The one parameter algebraic thawing form (solid black curve, 
Eq.~\ref{eq:alg}) accurately fits $w(z)$ at all redshifts, for the 
thawing models of linear, quadratic, quartic, and PNGB potentials. 
} 
\label{fig:alg9} 
\end{figure}

Figure~\ref{fig:alg9} demonstrates the fit of the algebraic form to the 
exact solutions for thawing $w(z)$, with $w_0=-0.9$. (Note that thawing 
models with greater present deviation from $-1$ are already in some 
tension with current data.) The agreement is better than 0.1\% 
in $w(z)$ (even better for the observables like distances and $H$) for the 
favored models of the linear potential and PNGB, and better than 0.2\% and 
0.3\% for the quadratic and quartic potentials.

While monomial potentials indeed have only one parameter to describe the 
equation of state, PNGB models have two: the steepness of the potential, 
given by the symmetry breaking scale $f$, and the initial field position. 
For steeper potentials, the algebraic form fit can degrade to $\approx1\%$ 
in $w$ (alternately we could adjust the values of $b$ or $p$ in the original 
algebraic form of Eq.~\ref{eq:algorig}, but there will still be a range 
of PNGB models that cannot be captured to much better than 1\% in $w(z)$ 
by only one parameter). 

Despite the success, it is important to note we are not trying to fit 
$w(z)$ per se (as it is not an observable), but rather to obtain a 
realistic enough representation that our constraints on distinguishing 
thawing models from $\Lambda$ in the next section are accurate. 
Table~\ref{tab:dev} demonstrates the success of both the algebraic thawer 
(Eq.~\ref{eq:alg}) and constrained \wowa\ form (Eq.~\ref{eq:wathaw}) in 
matching the observables. (Changing the value of, e.g., the matter density 
would slightly degrade the accuracy of fitting $w(z)$, but improve the 
accuracy 
of fitting $d$ or $H$ since there is more freedom allowed for the fit.) The 
accuracy is more than sufficient for next generation observations seeking to 
distinguish thawing dark energy from a cosmological constant $\Lambda$.

\begin{table}[!htb]
\begin{tabular}{l|ccc} 
Model\ \ & $\quad d(z)\quad$ & $\quad H(z)\quad$ & $\quad d_{\rm lss}\quad$ \\ 
\hline 
Linear potential & 0.06\% & 0.01\% & $10^{-5}$ \\ 
$\phi^2$ & 0.08\% & 0.02\% & $10^{-5}$ \\ 
$\phi^4$ & 0.16\% & 0.03\% & $10^{-5}$ \\ 
PNGB ($f/M_{\rm Pl}=2$) & 0.03\% & 0.01\% & $10^{-5}$ \\ 
Constrained $w_a$ & 0.09\% & 0.01\% & $4\times 10^{-5}$ \\ 
\hline 
$\Lambda$ & 3.5\% & 0.7\% & 0.5\% 
\end{tabular} 
\caption{Maximum deviations in the observables -- the comoving distance 
$d(z)$, Hubble parameter $H(z)$, and distance to CMB last scattering 
$d_{\rm lss}$ -- over all redshifts, relative to the algebraic thawing form, 
are given for the exact solutions of 
various thawing dark energy models with 
$w_0=-0.9$. The matter density and 
Hubble parameter are fixed ($\om=0.3$, $h=0.7$). 
} 
\label{tab:dev} 
\end{table}

\section{Constraining Thawers} \label{sec:fit} 

As Table~\ref{tab:dev} shows, the maximum difference between thawing dark 
energy deviating from the cosmological constant at late times to $w_0=-0.9$ 
is 3.5\% in distance (at $z=0.5$). However, this was with fixing all other 
cosmological parameters so in reality covariances make the distinction more 
challenging. 
Let us see if our two one-parameter thawing forms give data the leverage to 
distinguish such dynamical dark energy from a cosmological constant. 

To test these models with expansion history measurements we consider 
supernova distances and the CMB distance to last scattering. (We could use 
baryon acoustic oscillation distances, but supernovae are somewhat more 
sensitive to the equation of state, especially at the needed low redshift.) 
For supernovae
(SN), we consider a future sample of 150 SN at $z<0.1$, 900 between 
$z=0.1$--1, and 42 between $z=1$--1.7, with a magnitude systematic of 
$0.02(1+z)$. Given the systematic control out to $z=1$, this roughly 
represents a sample from ground based imaging surveys such as the Dark 
Energy Survey or Large Synoptic Survey Telescope (LSST). We also analyze 
an extended sample inspired by the DESIRE proposal \cite{desire} for  
the Euclid satellite, with 7200 SN between $z=0.1$--1 and 1000 at 
$z=1$--1.6, with the same systematics control out to $z=1.6$. These are 
simply rough estimates of potential data. For the CMB we use Planck quality 
distance to last scattering measurement at the 0.2\% level and a prior on the 
physical matter density $\Omega_m h^2$ of 0.9\%. 

The parameters for constraining the expansion history include the matter 
density $\om$, dynamical equation of state parameter $w_0$, supernova 
absolute magnitude $\mathcal{M}$, and Hubble constant $h$. 
Spatial flatness is assumed. 
Using the future data, the constraint on the equation of state parameter 
is $\sigma(w_0)\approx0.06$. If the thawing dark energy has reached 
$w_0=-0.9$ today (and a stronger deviation is already disfavored by present 
data), then this yields at most a $\approx1.7\sigma$ distinction from 
$\Lambda$. A stronger discriminant would be useful. 

Interestingly, the extended distance sample does not improve substantially 
the constraints here. While the baseline sample gives $\sigma(w_0)=0.063$ and 
0.062 for the algebraic thawer and calibrated $w_a$ forms, the extended 
sample gives 0.056 and 0.055. The reason for this insensitivity is that 
thawers tend to deviate appreciably from $\Lambda$ only at low redshift; 
for example thawers as represented by the algebraic form have $w(z=1)=-0.98$ 
while eventually reaching $w(z=0)=-0.9$. While supernova distances are indeed 
sensitive to the equation of state value at low redshift, they are not 
sensitive enough. 

The question then is what observational probe can more keenly test the 
low redshift equation of state. Recently, cosmic redshift drift has been 
recognized as highly sensitive at low redshift \cite{drift}. This constrains 
the quantity 
\be 
\dot z=(1+z)\,H_0-H(z) \ . 
\ee 
For the (general) algebraic thawer the Hubble parameter can be written 
analytically as  
\be 
\left(\frac{H(z)}{H_0}\right)^2=\om a^{-3}+(1-\om)\, 
e^{\frac{3(1+w_0)}{\alpha p}\left[1-(\alpha a^3+\beta)^{p/3}\right]} \, 
\ee 
where $\alpha=1/(1+b)$, $\beta=b/(1+b)$. 
Since redshift drift is wholly unproven, we conjecture a single, modest 5\% 
measurement at redshift $z=0.5$. This one data point improves the distinction 
between thawers and $\Lambda$ by a factor of two, reducing $\sigma(w_0)$ 
to 0.030, marginalized over the other cosmological parameters, 
and hence a $3.3\sigma$ distinction from a cosmological constant. 

Of course as $w_0\to-1$ the dark energy 
would be increasingly difficult to distinguish from $\Lambda$. 
Section~\ref{sec:field} discusses particle physics implications of such 
near-$\Lambda$ fields.

\section{Fine Tuned Fields } \label{sec:field} 

Thawing fields have a limit in which they are identical to a cosmological 
constant, i.e.\ when $w_0=-1$, and so the full class can never be ruled 
out if the data are consistent with $\Lambda$. We examine whether fields 
in such a limit are particularly fine tuned or disfavored in particle 
physics terms. 

\begin{figure}[htbp!]
\includegraphics[width=\columnwidth]{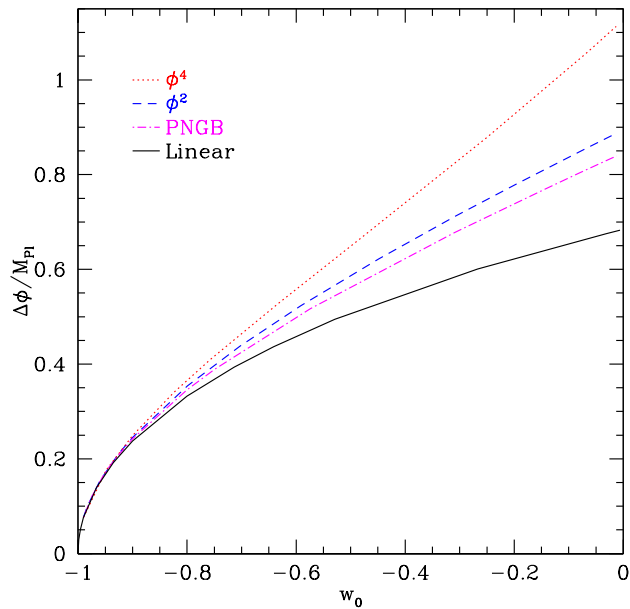} 
\caption{The distance the field rolls up to the present is plotted vs 
the present equation of state for several thawer models. The field only 
traverses a short distance for, say, $w_0=-0.9$ but must start several 
$M_{\rm Pl}$ from the potential minimum. 
} 
\label{fig:dphi} 
\end{figure}

First consider quadratic and quartic potentials as thawers. 
Like all thawers, the fields do not roll very far during their evolution 
up to the present, as shown in Fig.~\ref{fig:dphi}. 
However, to attain even only 
$w_0=-0.9$ today, the fields must start $2.6\,M_{\rm Pl}$ and 
$5.1\,M_{\rm Pl}$ respectively from the minima. The potentials must 
be sufficiently steep locally to be able to reach $\Omega_{de,0}=0.7$ 
today once they are released from Hubble friction to roll. Values of 
$w_0$ closer to $-1$ exacerbate the situation, with the fields needing 
to start at $7.9\,M_{\rm Pl}$ and $15.7\,M_{\rm Pl}$ respectively to 
reach, say, $w_0=-0.99$ and hence be easily confused with $\Lambda$. 
Especially since such potentials have no protection against radiative 
corrections, we might be concerned about such superPlanckian values 
(despite the field rolling over less than a Planck mass). If we restrict 
the field initial position to lie within one Planck mass of the minimum, 
these potentials cannot deliver dark energy density 
$\Omega_{de}>0.45$ and 0.17 respectively.

The linear potential, however, is robust. For $V(\phi)=V_0(1+\alpha\phi)$, 
while the slope of the potential $\alpha V_0\approx 10^{-120}$ as for any 
quintessence, the parameter $\alpha$ itself is of order unity. For 
$w_0=-0.9$ we have $\alpha=0.72$, and behavior close to $\Lambda$ can be 
obtained without fine tuning $\alpha$, e.g.\ $w_0=-0.99$ comes from 
$\alpha=0.25$. Furthermore the linear potential is substantially protected 
against quantum corrections. In the future it leads to a cosmic doomsday 
\cite{doom}, actually essential for the recent solution of the cosmological 
constant problem known as the sequester \cite{sequester}. Short of doomsday, 
the linear potential can be extremely difficult to distinguish from 
$\Lambda$, however, and so upcoming observations will not be able to rule 
out fully the thawing class. 

For the PNGB model, a shift symmetry protects the potential form against 
quantum corrections. Nevertheless, one might be concerned about the 
physics validity if the symmetry energy scale $f>M_{\rm Pl}$. As $f$ 
decreases, and the potential steepens, it becomes harder to attain 
$\Omega_{de,0}=0.7$. The field must be fine tuned to an initial position 
$\phi_i$ closer and closer to the potential maximum, out of all the 
range $[0,\pi f]$. For fixed $\phi_i/f$, $\Omega_{de,\rm{max}}\propto f^2$, 
while for fixed $\Omega_{de,0}$, $(\phi_i/f)_{\rm max}\sim e^{-1/f}$ 
\cite{calde}. Thus small $f$ is problematic. To obtain $w_0<-0.99$ 
with $f=0.5\,M_{\rm Pl}$ requires $\phi_i/f<0.18$, i.e.\ only allowing 6\% 
of the full range, while for $f=0.2\,M_{\rm Pl}$ a 0.5\% fine tuning 
is required. Still, a reasonable region near $f/M_{\rm Pl}\approx1$ is 
allowed and can give an expansion history effectively indistinguishable 
from $\Lambda$. 

Thus we see that nothing prevents thawing fields, and in particular the 
two technically natural and hence robust models of the linear potential 
and PNGB, from approaching $\Lambda$ beyond the ability of upcoming  
experiments to distinguish.

\section{Conclusions} \label{sec:concl} 

Great strides have been made in constraining dark energy physics with 
ever improving observations. Freezing dark energy, including early dark 
energy, is substantially limited by the data as an explanation for cosmic 
acceleration. Probes of the expansion history will continue to become more 
incisive, and probes of the growth of large scale structure will impose 
tighter constraints on more elaborate theories involving dark energy 
clustering, coupling, and modified gravity. 
However even if the data continue to be consistent with a cosmological 
constant, quintessence remains a viable option in its thawing class. 

The standard dark energy equation of state parametrization 
$w(a)=w_0+w_a(1-a)$ is an excellent global approximation over a wide 
range of freezer, thawed, or modified gravity models, to the 0.1\% level in 
the observables. A constrained, one parameter form with 
$w_a=-1.58(1+w_0)$ can be used to model specifically the thawing class. 
Another option is to use the algebraic thawer form of Eq.~(\ref{eq:alg}) that 
follows the thawing physics to match $w(z)$ as well as the observables. 
We use these not as fitting forms per se, but to represent the thawing class 
as a whole for comparison to a cosmological constant, to evaluate the 
discriminatory leverage of forthcoming data. 

We find that next generation distance data can reveal modest, but consistent 
distinctions of thawers from $\Lambda$ if $w_0\gtrsim -0.9$. The leverage 
in distinguishing the physics doubles with inclusion of a single, 5\% 
measurement by the prospective probe of cosmic redshift drift. This gives 
further motivation for its exploration and development. In particular, since 
the signature of the thawing deviation increases at very low redshift -- 
e.g.\ the deviation is five times greater at $z=0$ than at $z=1$ -- then the 
recently identified power of redshift drift at low redshift \cite{drift} is 
especially valuable. 

While quintessence in the form of thawing fields cannot be ruled out even if 
the data become increasingly consistent with a cosmological constant, despite 
the very different physics, such behavior does point to preferred models. Both 
the linear potential and PNGB are technically natural, easing high energy 
quantum effects by their shift symmetry. Indeed the linear potential has 
recently been found essential in one method for solving the original 
cosmological constant problem \cite{sequester}. If future data show some 
definite signature in expansion or growth away from $\Lambda$, we will be 
able to focus our efforts on specific physical properties. Even if $\Lambda$ 
continues to be a good fit, we will be drawn either to explaining its 
magnitude or to finding the informative high energy physics origins of these robust 
thawing quintessence potentials.

\acknowledgments 

This work has been supported by DOE grant DE-SC-0007867 and the Director, 
Office of Science, Office of High Energy Physics, of the U.S.\ Department 
of Energy under Contract No.\ DE-AC02-05CH11231. 


\end{document}